\documentclass[english]{emulateapj}
\usepackage{lmodern}

\usepackage[T1]{fontenc}
\usepackage[latin9]{inputenc}
\usepackage{xcolor}
\usepackage{array}
\usepackage{float}
\usepackage{booktabs}
\usepackage{multirow}
\usepackage{amssymb}
\usepackage{graphicx}
\usepackage{esint}
\PassOptionsToPackage{normalem}{ulem}
\usepackage{ulem}

\makeatletter

\providecommand{\tabularnewline}{\\}
\providecolor{lyxadded}{rgb}{0,0,1}
\providecolor{lyxdeleted}{rgb}{1,0,0}

\makeatother

\usepackage{babel}
\begin{document}

\title{Formation and evolution of nuclear star clusters with in-situ star-formation:\\
Nuclear cores and age segregation }

\author{Danor Aharon \& Hagai B. Perets}

\affil{Physics Department, Technion - Israel Institute of Technology, Haifa,
Israel 32000}
\begin{abstract}
Nuclear stellar cluster (NSCs) are known to exist around massive black
holes (MBHs) in galactic nuclei. Two formation scenarios were suggested
for their origin: (1) Build-up of NSCs from consecutive infall of
stellar cluster and (2) Continuous in-situ star-formation. Though
the cluster-infall scenario has been extensively studied, the in-situ
formation scenario have been hardly explored. Here we use Fokker-Planck
(FP) calculations to study the effects of star formation on the build-up
of NSCs and its implications for their long term evolution and their
resulting structure. We use the FP equation to describe the evolution
of stellar populations, and add appropriate source terms to account
for the effects of newly formed stars. We show that continuous star-formation
even 1-2 pc away from the MBH can lead to the build-up of an NSC with
properties similar to those of the Milky-way NSC. We find that the
structure of the old stellar population in the NSC with in-situ star-formation
could be very similar to the steady-state Bahcall-Wolf cuspy structure.
However, its younger population do not yet achieve a steady state.
In particular, formed/evolved NSCs with in-situ star-formation contain
differential age-segregated stellar populations which are not yet
fully mixed. Younger stellar populations formed in the outer regions
of the NSC have a cuspy structure towards the NSC outskirts, while
showing a core-like distribution inwards; with younger populations
having larger core sizes. In principal, such a structure can give
rise to an apparent core-like radial distribution of younger stars,
as observed in the Galactic Center. 
\end{abstract}

\section{INTRODUCTION}

Nuclear stellar clusters, (NSCs) hosting massive black holes (MBHs)
are thought to exist in a significant fraction of all galactic nuclei.
Their origin is still not well understood. Two main scenarios were
suggested for their origin: (1) The cluster infall scenario, in which
stellar clusters inspiral to the galactic nucleus, disrupted, and
thereby build up the nuclear cluster \citep{1975ApJ...196..407T,1993ApJ...415..616C,2011ApJ...729...35A,2013ApJ...763...62A}.
(2) The nuclear star formation (SF) scenario, in which gas infalls
into the nucleus and then transforms into stars through star formation
processes \citep{1982A&A...105..342L}. Here we focus on the latter
process, and study the long term effects of SF on the formation and
evolution of NSCs. 

The structure, evolution and dynamics of NSCs have been extensively
studied in recent years. These explored the general dynamics of NSCs,
and in particular NSCs similar to the well-observed NSC in the Milky
Way Galactic Center (GC). The presence of a young stellar disk in
the central pc of GC, as well as dense concentration of HII regions
and young stars throughout the central 100 pc of the Milky-way \citep{2004ApJ...601..319F}
provide evidence for a continuous star-formation in this region, both
today and in the past \citep{2010RvMP...82.3121G}. The majority of
the observed stars in the GC are either late type main sequence stars
or red giants, suggesting that most of the stars in the GC likely
formed at least a few Gyr ago (\citet{2010RvMP...82.3121G}). \citet{2011ApJ...741..108P}
argued that the star formation rate in the central pc of the GC has
decreased since then until it has dropped to a deep minimum 1-2 Gyrs
ago and increased again during the last few hundred Myrs, suggesting
that there are number of epochs of star formation. Evidence for star
formation exists in other extagalactic NSCs \citep[e.g. ][and references therein]{2006AJ....132.2539S,2014MNRAS.441.3570G}.
\citet{2005ApJ...618..237W} argued that NSCs are protobulges that
grow by repeated accretion of gas and subsequent star formation, \citet{2006ApJ...650L..37M}
suggested a NSC in-situ star formation model regulated by momentum
feedback. 

These various studies provide further motivation and suggest that
star-formation has an important role in shaping NSCs and their evolution.
Nevertheless, very few studies systematically explored its role, while
most studies focused on the cluster-infall scenario. Here, we focus
on the role of in-situ SF in NSCs, and explore its implication both
for the build-up of the NSC, as well as the long term evolution and
structure of NSCs. Our work makes use of the Fokker-Planck diffusion
equations, first used by \citet{1976ApJ...209..214B} in this context,
to describe the dynamics of stellar populations in dense clusters
around MBHs.

In the following we begin by a brief description of the original Fokker-Planck
method used to study MBH-hosting clusters, and then describe our approach
of adding an additional source term and multiple populations to account
for the effects of SF. We than present the resulting NSC structure
arising from the SF-included evolution, as characterized by the number
density and age distribution of its constituent stellar populations.
We explore various possible models for the NSC SF history and study
their outcomes. Finally, we discuss our results and their implications
for the formation of NSC, nuclear cores and age segregated populations
and summarize.

\section{Relaxation and evolution of NSCs around MBHs}

NSCs are complex interacting systems. Their evolution and dynamics
are mainly affected by 2-body relaxation (see section \ref{sub:The-basic-model}).
Theoretical studies of NSCs based on 2-body relaxation include simulations
of many-particle systems under a potential generated by a massive
body (MBH). One method to obtain realistic model of NSCs is through
a full N-body simulation. However, running these simulations to describe
the evolution of realistic NSCs is currently still too computationally
expensive, and only a few studies involving N-body simulations of
relatively small NSCs have been done (e.g. \citealp{2004ApJ...613L.109P,2012ApJ...750..111A}
and references therein). Other, more efficient methods, though more
limited in their application and dependence on specific assumptions,
make use of the Fokker-Planck (FP) diffusion equation approach \citep{1976ApJ...209..214B,1978ApJ...226.1087C}
(also solved through Monte-Carlo simulations \citealt{1978ApJ...225..603S,2006ApJ...649...91F}).
Here we follow the latter and study the evolution of NSCs by numerically
solving the FP equations following the approach first used by Bahcall
\& Wolf (BW;1976,1977) in this context. However, we supplement the
basic equations, for the first time, with a source term accounting
for SF, as well as use a large number of distinct stellar populations
to account for different SF epochs. We also parallelized the FP code
as to allow us to easily explore the NSC with both high resolution
and a large number of stellar population. With these tools in hand
we obtain for a model for the long term evolution of SF in NSCs with
in-situ SF. In the following we provide a brief discussion of the
original FP approach, and then present the additional SF source term,
as well as discuss the model assumptions and limitations.

\subsection{Fokker-Planck analysis}

\subsubsection{The basic model\label{sub:The-basic-model}}

Following the model introduced by Bahcall \& Wolf (1976, 1977), We
simulate the evolution in time, $t,$ of the energy, $E,$ distribution
function (DF) - $f(E,t)$ and the number density of stars in a spherical
system around a MBH, with a mass of $M_{\bullet}=4\times10^{6}M_{\odot}$
(the latter is chosen to be similar to the MBH mass in the GC). We
focus on the distribution of stars in central few pcs, and in particular
in the range between the tidal radius - $r_{t}\approx r_{*}(M_{\bullet}/M_{*})^{1/3}$below
which regular stars are disrupted by the MBH, and the radius of influence
where the stellar motions are dominated by the MBH potential, defined
by $r_{h}=GM_{\bullet}/\sigma_{\star}^{2}$ , where $r_{*}$, and
$M_{*}$ are the typical radius and mass of stars in the NSC, respectively;
and $\sigma_{\star}$ is the velocity dispersion of stars just outside
the NSC. The velocity dispersion in our model is\textbf{ $\sigma\thickapprox75km/sec$
}(consistent with M-Sigma relation;\textbf{ }\citet{Ferrarese2000}).
For $r>r_{h}$ the original BW model assumes the existence of a ``thermal
bath'' which supplies stars to the inner region the galactic nucleus.
The stellar orbits are assumed to be Keplerian in this range. The
model timescale is dominated by the relaxation time, which is defined
for a single mass system as
\begin{equation}
T_{r}\approx\frac{\sigma^{3}}{G^{2}M_{*}\rho\ln(\Lambda)}\label{eq:relaxation}
\end{equation}
Where $\rho$ is the stellar density, and $\ln(\Lambda)$ is the Coulomb
logarithm, a factor which is related to the scale of the system ($\ln(\Lambda)\approx10)$.
The simulation is done by solving the time and energy-dependent, angular
momentum-averaged particle conservation equation. This FP equations
describes the two-body diffusion of stars in energy phase space within
the boundary conditions, i.e. between the fixed thermal bath reservoir
of stars stars outside the radius of influence and the MBH sink, where
the distribution is null on the tidal radius. The stars in the reservoir
($r>r_{h})$ are assumed to have a Maxwellian distribution, with the
energy dependent distribution:

\begin{equation}
f(E,t)=n_{0}\left(2\pi\sigma^{2}\right)^{-\frac{3}{2}}e^{\frac{E}{\sigma^{2}}},
\end{equation}
where $n_{0}$ is the number density of stars in the cluster\textquoteright{}s
core. Inside the radius of influence ($r<r_{h})$ the distribution
function is evaluated using the FP diffusion equation with a loss-cone
term (which will be discussed in the following):

\begin{equation}
\frac{\partial f(E,t)}{\partial t}=-AE^{-\frac{5}{2}}\frac{\partial F}{\partial E}-F_{LC}(E,T)
\end{equation}

where 
\begin{equation}
A=\frac{32\pi^{2}}{3}G^{2}M_{*}^{2}\ln(\Lambda).
\end{equation}

The term $F=F[f(E),E]$ is related to the stellar flow, and plays
an important role in the evolution of the stellar cluster. It presents
the flow of stars in energy space due to two-body relaxation, it is
defined by:

\begin{equation}
F=\int dE'\left(f(E,t)\frac{\partial f(E',t)}{\partial E'}-f(E',t)\frac{\partial f(E,t)}{\partial E}\right)\left(\max(E,E')\right)^{-\frac{3}{2}}.\label{eq:flow_eq}
\end{equation}

The sink term $F_{LC}(E,T)$ is the rate at which stars, with energies
in the interval $(E,E+dE)$, flow into the MBH and are therefore lost
from the system (hence the term loss-cone) through disruption by the
MBH at the tidal radius, and it is approximated by

\begin{equation}
F_{LC}^{Empty}(E,T)\propto\frac{f(E,t)}{T_{R}}.
\end{equation}

The added $F_{LC}(E,T)$ term corresponds to the empty loss-cone regime;
once the rate of angular momentum diffusion exceeds the ``pinhole\textquotedblright{}
regime limit, the full loss cone regime dominates the tidal disruption
dynamics and the tidal disruption rate saturates. We therefore take
$F_{{\rm LC}}(E,T)=min(F_{LC}^{Empty},\, F_{LC}^{Eull}(E,T))$, as
a simple approximation accounting for the transition between the empty
and full loss-cone regimes (see \citealp{1977ApJ...211..244L,1977ApJ...215...36Y,2007ApJ...656..709P}).
Note that the BW 1D FP equation allows stars to diffuse only in energy
phase space, and do not account for stars with small angular momentum
that may be disrupted as they approach the MBH though highly eccentric
orbits, which pericenter becomes smaller than the tidal radius. The
above loss-cone term is an added term which effectively accounts for
the loss through angular momentum \citep{1976MNRAS.176..633F,1977ApJ...216..883B},
which is not properly accounted for in the 1D FP equation in energy
phase space.

\subsubsection{The star-formation source term}

In our model we simulate the star formation in the GC through adding
an extra source term component to the Fokker-Planck equation.\textbf{
}\citet{2015MNRAS.446..710H} also used a source term with FP equation
for modeling inflow of planetesimals. The term's value and range in
our model are determined according to the number of new stars added
in the appropriate region. We simulate multiple stellar populations
forming at different epochs, and follow the evolution of their distribution.
The initial conditions of the simulated NSCs include a background
stellar population that either evolves from an initial steady state
BW distribution or is initially zero and is then entirely built up
from the in-situ star formation. 

The modified Fokker-Planck equation with the addition of the source
term (and the loss-cone term) has the form:

\begin{equation}
\frac{\partial f(E,t)}{\partial t}=-AE^{-\frac{5}{2}}\frac{\partial F}{\partial E}-F_{LC}(E,T)+F_{SF}
\end{equation}

\begin{equation}
F_{SF}=\frac{\partial}{\partial t}\left(\Pi(E)E_{0}E^{\alpha}\right)\label{eq:source term}
\end{equation}

$\Pi(E)$ is a rectangular function, which boundaries correspond to
the region where new stars are assumed to from; $E_{0}$ is the source
term amplitude; and $F_{SF}$ is a power-law function with a slope
$\alpha$, defining the SF distribution in phase space . We simulated
a number of NSC evolutionary scenarios, taking different models for
the SF function (rate and spatial structure) and for the background
population (see section \ref{sec:models}). The chosen slope of the
SF function was motivated by the observed power-law \citep{2009ApJ...703.1323D,2009ApJ...697.1741B}
distribution of young stars observed in the young stellar disk in
the GC.

\subsection{Methods}

In order to calculate the distribution and the number density profile
of stellar populations in the modeled NSCs, we numerically integrate
the differential equations presented in BW77 (including an effective
tidal disruption loss-cone sink term), supplemented with the additional
source, in order to follow the long term evolution of the NSC stellar
populations. Our code is based on the original code written by \citet{2006ApJ...645L.133H},
which was significantly modified, as follows: (1) It includes the
added SF source term as discussed above; (2) It now includes both
empty and full loss-cone dynamics, which were not consistently included
before; (3) we use the simple moving window average method to deal
with numerical instabilities, which lead to fluctuations in the numerical
solution in previous treatments; and (4) to improve the numerical
resolution and reduce computation time, our algorithm was parallelized
using OpenMP library, allowing for efficient calculation of higher
resolution grids and the modeling of a large number of stellar populations.

\section{Models}

\label{sec:models}

\subsection{Basic assumptions and limitations}

In our work we use a simplified 1D FP model, and consider only a single
mass for all stars. Multi-mass populations can affect the dynamical
evolution of the NSC both through the potentially faster relaxation
induced by more massive stars, as well as through mass-segregation
processes, which may induce different distributions for different
mas stars (e.g. \citet{1977ApJ...216..883B,2009ApJ...697.1861A}).\textcolor{red}{{}
}\textcolor{black}{The first effect can be effectively studied to
some extent even in single-mass populations by changing the mass of
all stars, treating it as the mean mass; we therefore use models with
stellar masses of either ${\rm 0.6M_{\odot}}$or ${\rm 0.8M_{\odot}}$.
The latter effect of mass-segregation requires more complex models.
Note that the typical present day mass-function has a very steep power
law, therefore the low frequency of more massive stars has relatively
little effect on the relaxation processes, even given the relaxation
time dependence on the stellar masses. Moreover, the vast majority
of stars are of comparable mass to the mean masses of used in the
single-mass approximation (massive stars become low mass compact objects
on timescales typically much shorter than the relaxation time), and
therefore mass-segregation effects are weak. The most significant
effect is for stellar black holes, whose mass is considerably larger
than the typical stellar mass. These objects are therefore likely
to segregate to the inner regions, potentially contributing to production
of extreme mass-ratio gravitational wave inspirals, and slightly accelerate
relaxation processes in the inner-most region of the cusp}\textbf{\textcolor{black}{{}
}}\textcolor{black}{\citep[e.g.][]{2006ApJ...649...91F,2006ApJ...645L.133H}.}\textbf{\textcolor{black}{{}
}}\textcolor{black}{We do not model these latter effects;} multiple
mass populations and mass-changes due to stellar evolution (e.g. \citealt{1991ApJ...370...60M}),
are beyond the scope of this study and will be explored elsewhere.
We consider only NSCs hosting a MBHs of $4\times10^{6}$ ${\rm M_{\odot}}$
similar to the GC MBH, and assume no growth of the MBH during the
evolution.

Similar to previous studies \citep{1976ApJ...209..214B,2006ApJ...645L.133H},
it is assumed that the gravitational potential is always dominated
by the MBH, and the contribution of the NSC stellar mass to the gravitational
potential is neglected. This assumption is reasonable as long as the
NSC mass is smaller, or comparable to that of the MBH. This is generally
true for our simulated NSC, as we consider regions of up to a few
pcs away from the MBH; typically comparable to the MBH radius of influence.
We do note that in the simulations with the highest stellar densities
this assumption becomes somewhat weaker for the outskirts of the NSC,
a few pcs away from the MBH. Nevertheless, the range at which our
analysis exceeds the $r_{h}$ is not more than twice of it, and the
difference in the typical parameters (i.e. $T_{R}$, $r_{h}$, $\sigma$)
is expected to be small. Moreover, the relaxation times in these regions
are very long and very little evolution, if any, is expected to occur
there as to significantly affect the NSC bulk population. 

Newly formed stars are assumed to form isotropically. Hence, we do
not self-consistently take into account the possibility of star-formation
in a disk-like configuration (e.g. the GC stellar disk ;\citealp{2003ApJ...590L..33L,2009ApJ...697.1741B,2009ApJ...690.1463L,2009MNRAS.394..191H,2012MNRAS.427.1793G},
and stellar disks in extra-galactic NSCs; \citet{2006AJ....132.2539S}).
In principle other source terms of stars can exist, such as stars
captured following a binary disruption \citep[e.g.][]{1988Natur.331..687H,2003ApJ...592..935G,2007ApJ...656..709P},
which could have very different distribution and affect the NSC dynamics
\citep{2009ApJ...702..884P,2010ApJ...719..220P}. We postpone treatment
of such qualitatively different source terms to future studies.

Finally, our models only include 2-body scattering relaxation processes
by stars, and do not include effects of coherent resonant relaxation
\citep{1996NewA....1..149R,2006ApJ...645.1152H} or 2-body relaxation
by massive perturbers \citep{2007ApJ...656..709P}. 

We consider two general types of models. In the first we study the
build-up of an NSC around a MBH, in which all stars originate from
long-term in-situ SF processes, modeled through the FP SF source term.
In the second, we consider a pre-existing NSC with an initial steady-state
BW structure, and study the effect of adding SF to the cluster, to
explore the distribution and evolution of stellar populations which
form at different epochs.

\subsection{Model parameters: Initial NSC structure, stellar components and star
formation histories}

We simulated various scenarios that describe the evolution of NSCs,
and the stellar populations that compose them. We consider a variety
of different scenarios, summarized in Tables \ref{tab:Initial-structures,-SF}
and \ref{tab:results}. The specific initial conditions, and the characteristics
of SF in NSCs are not known, and we therefore study a variety of possible
cases. These differ in their SF rates and history, the regions where
SF occur, the typical masses of stars in the NSCs, and whether they
include a pre-existing NSC or not. Naturally, the different scenarios
are not exhaustive, but they provide a basic characterization of a
range of plausible, though simplified scenarios for NSC evolution.
The regions where SF occurs are motivated by the observed young stars
in the young stellar disk (for models with SF in the inner $0.5$
pc), and the circum-nuclear disk in the GC as well as evidence for
long term SF in the central 100 pc of the MW (for models with SF in
the inner 1.5-5 pc region; e.g. \citealt{2004ApJ...601..319F,2011ApJ...732..120O}).
We consider only single mass models; in particular we study cases
where the mean mass of stars is assumed to be either 0.8 or 0.6 $M_{\odot}$.
The stellar radii are also changed, assuming a mass-radius relation
of $R=(M/M_{\odot})^{0.8}R{}_{\odot}$. The corresponding tidal radii
used are then changed accordingly; see table\textbf{ \ref{tab:Initial-structures,-SF}}.

All the models are evolved for a Hubble time. In most models one population
represents the background initial stellar population of the cusp (or
the stellar population that formed at an early stage of the NSC evolution)
and ten additional populations are later introduced consecutively
to study the evolution of stars formed at different epochs, up to
1 (or 3) Gyrs ago. Each population represents stars forming over a
continuous range of 100 (or 300) Myrs, and introduced continuously
after the the end of the previous SF epoch (beside model 8, where
we consider 100 Myrs SF bursts once every Gyr, i.e. separated by long
quiescent times). Stellar populations formed much earlier typically
have sufficient time to achieve an almost steady state solution, very
similar to the BW solution, and are not studied in details by themselves,
but only considered together (as part of the first/background population),
while younger populations are not relaxed, and show a differential
behavior (see below).
/newpage
\begin{table*}
\begin{tabular}{|>{\centering}p{1cm}|>{\centering}p{8.5cm}|c|>{\centering}p{3.5cm}|}
\hline 
 \# & Initial NSC structure and SF history  & Stellar mass & Tidal radius\textbf{ }($pc$)\tabularnewline
\hline 
\hline 
i & {\footnotesize{Background BW population + 10 populations that form
in the last 1 Gyr. }} & {\footnotesize{$0.8M_{\odot}$}} & {\footnotesize{$3.2\times10^{-6}$}}\tabularnewline
\hline 
ii & {\footnotesize{Background BW population + 10 populations that form
in between 4 Gyr and 5 Gyr. }} & {\footnotesize{$0.8M_{\odot}$}} & {\footnotesize{$3.2\times10^{-6}$}}\tabularnewline
\hline 
iii & {\footnotesize{Background population }}\textbf{\footnotesize{from
SF}}{\footnotesize{ + 10 populations that form in the last 1 Gyr. }} & {\footnotesize{$0.8M_{\odot}$}} & {\footnotesize{$3.2\times10^{-6}$}}\tabularnewline
\hline 
iv & {\footnotesize{Background BW population + 10 populations non continuously
form in the last 5 Gyr. }} & {\footnotesize{$0.8M_{\odot}$}} & {\footnotesize{$3.2\times10^{-6}$}}\tabularnewline
\hline 
v & {\footnotesize{Background BW population + 10 populations that form
in the last 3 Gyr. }} & {\footnotesize{$0.8M_{\odot}$}} & {\footnotesize{$3.2\times10^{-6}$}}\tabularnewline
\hline 
vi & {\footnotesize{Background population }}\textbf{\footnotesize{from
SF}}{\footnotesize{ + 10 populations that form in the last 3 Gyr. }} & {\footnotesize{$0.8M_{\odot}$}} & {\footnotesize{$3.2\times10^{-6}$}}\tabularnewline
\hline 
vii & {\footnotesize{Background BW population + 10 populations that form
in the last 3 Gyr. }} & {\footnotesize{$0.6M_{\odot}$}} & {\footnotesize{$2.8\times10^{-6}$}}\tabularnewline
\hline 
viii & {\footnotesize{Background population }}\textbf{\footnotesize{from
SF}}{\footnotesize{ + 10 populations that form in the last 3 Gyr.}} & {\footnotesize{$0.6M_{\odot}$}} & {\footnotesize{$2.8\times10^{-6}$}}\tabularnewline
\hline 
\end{tabular}\label{tab:scenarios}\caption{\label{tab:Initial-structures,-SF}Initial structures, SF histories
and stellar masses for the NSC models. }
\end{table*}

\begin{table*}
\begin{tabular}{>{\centering}p{0.7cm}>{\centering}p{1cm}>{\centering}p{1.6cm}>{\centering}p{2.4cm}>{\centering}p{1.48cm}>{\centering}p{1.5cm}>{\centering}p{0.15cm}>{\centering}p{1.6cm}>{\centering}p{1.6cm}>{\centering}p{1.75cm}}
\toprule 
\multicolumn{6}{c}{{\large{Input parameters}}} &  & \multicolumn{3}{c}{{\large{Results}}}\tabularnewline
\midrule
\midrule 
{\scriptsize{Pop}} & {\scriptsize{model}} & {\scriptsize{SF rate ($yr^{-1}$)}} & {\scriptsize{SF duration for each population ($yr$) }} & {\scriptsize{Power-law}} & {\scriptsize{SF Range ($pc$)}} &  & {\scriptsize{Core size ($pc$)}} & {\scriptsize{Total stars}} & {\scriptsize{Relaxation time (Gyr)}}\tabularnewline
\midrule
\midrule 
\multirow{5}{0.7cm}{{\scriptsize{i}}} & {\scriptsize{1}} & {\scriptsize{$1\times10^{-4}$}} & {\scriptsize{$1\times10^{8}$ }} & {\scriptsize{-2}} & {\scriptsize{$2.5-3.7$}} &  & {\scriptsize{$0.1-2.1$}} & {\scriptsize{$5.2\times10^{6}$ }} & {\scriptsize{$8.1$}}\tabularnewline
\cmidrule{2-10} 
 & {\scriptsize{2a}} & {\scriptsize{$1\times10^{-4}$}} & {\scriptsize{$1\times10^{8}$ }} & {\scriptsize{-2}} & {\scriptsize{$0.05-0.1$}} &  & {\scriptsize{$0.002-0.05$}} & {\scriptsize{$5.3\times10^{6}$}} & {\scriptsize{$6.3$}}\tabularnewline
\cmidrule{2-10} 
 & {\scriptsize{2b}} & {\scriptsize{$1\times10^{-4}$}} & {\scriptsize{$1\times10^{8}$ }} & {\scriptsize{-2}} & {\scriptsize{$0.05-0.5$}} &  & {\scriptsize{$0.002-0.03$}} & {\scriptsize{$5.2\times10^{6}$}} & {\scriptsize{$7.0$}}\tabularnewline
\cmidrule{2-10} 
 & {\scriptsize{3}} & {\scriptsize{$1\times10^{-4}$}} & {\scriptsize{$1\times10^{8}$ }} & {\scriptsize{-2}} & {\scriptsize{$1.7-2.2$}} &  & {\scriptsize{$0.1-1.3$}} & {\scriptsize{$5.2\times10^{6}$}} & {\scriptsize{$8.1$}}\tabularnewline
\cmidrule{2-10} 
 & {\scriptsize{4}} & {\scriptsize{$1\times10^{-4}$}} & {\scriptsize{$1\times10^{8}$ }} & {\scriptsize{-2}} & {\scriptsize{$2.5-7.5$}} &  & {\scriptsize{$0.1-1.3$}} & {\scriptsize{$5.2\times10^{6}$}} & {\scriptsize{$8.1$}}\tabularnewline
\midrule
\midrule 
\multirow{2}{0.7cm}{{\scriptsize{ii}}} & {\scriptsize{5}} & {\scriptsize{$1\times10^{-4}$}} & {\scriptsize{$1\times10^{8}$ }} & {\scriptsize{-2}} & {\scriptsize{$2.5-7.5$}} &  & {\scriptsize{$0.001$}} & {\scriptsize{$5.1\times10^{6}$}} & {\scriptsize{$8.1$}}\tabularnewline
\cmidrule{2-10} 
 & {\scriptsize{6}} & {\scriptsize{$1\times10^{-4}$}} & {\scriptsize{$1\times10^{8}$ }} & {\scriptsize{-2.5}} & {\scriptsize{$2.5-3.7$}} &  & {\scriptsize{$0.001$}} & {\scriptsize{$5.2\times10^{6}$}} & {\scriptsize{$8.0$}}\tabularnewline
\midrule
\midrule 
{\scriptsize{iii}} & {\scriptsize{7}} & {\scriptsize{$1\times10^{-4}$}} & {\scriptsize{$1\times10^{8}$ }} & {\scriptsize{-2}} & {\scriptsize{$2.5-3.7$}} &  & {\scriptsize{$0-2.5$}} & {\scriptsize{$2.7\times10^{7}$}} & {\scriptsize{$2.6$}}\tabularnewline
\midrule
\midrule 
{\scriptsize{iv}} & {\scriptsize{8}} & {\scriptsize{$2\times10^{-5}$}} & {\scriptsize{$1\times10^{8}$ for each population for one interval
of formation. $9\times10^{8}$ break-interval}} & {\scriptsize{-2}} & {\scriptsize{$2.5-3.7$}} &  & {\scriptsize{$0.1$}} & {\scriptsize{$5.1\times10^{6}$}} & {\scriptsize{$9.8$}}\tabularnewline
\midrule
\midrule 
{\scriptsize{v}} & {\scriptsize{9}} & {\scriptsize{$1\times10^{-4}$}} & {\scriptsize{$3\times10^{8}$ for each population}} & {\scriptsize{-2}} & {\scriptsize{$2.5-3.7$}} &  & {\scriptsize{$0.003-1$}} & {\scriptsize{$5.2\times10^{6}$}} & {\scriptsize{$8.0$}}\tabularnewline
\midrule
\midrule 
{\scriptsize{vi}} & {\scriptsize{10}} & {\scriptsize{$1\times10^{-4}$}} & {\scriptsize{$3\times10^{8}$ }} & {\scriptsize{-2}} & {\scriptsize{$2.5-3.7$}} &  & no core & {\scriptsize{$1.3\times10^{7}$}} & {\scriptsize{$2.6$}}\tabularnewline
\midrule
\midrule 
{\scriptsize{vii}} & {\scriptsize{11}} & {\scriptsize{$1\times10^{-4}$}} & {\scriptsize{$3\times10^{8}$ }} & {\scriptsize{-2}} & {\scriptsize{$2.5-3.7$}} &  & {\scriptsize{$0.1-2.5$}} & {\scriptsize{$6.1\times10^{6}$}} & {\scriptsize{$12$}}\tabularnewline
\midrule
\midrule 
{\scriptsize{viii}} & {\scriptsize{12}} & {\scriptsize{$1\times10^{-4}$}} & {\scriptsize{$3\times10^{8}$ }} & {\scriptsize{-2}} & {\scriptsize{$2.5-3.7$}} &  & {\scriptsize{$0.01-1.5$}} & {\scriptsize{$2.5\times10^{7}$}} & {\scriptsize{$18$}}\tabularnewline
\bottomrule
\end{tabular}

\caption{\label{tab:results} SF properties for the NSC models, and their characteristic
properties after 10 Gyrs of evolution. }
\end{table*}

\section{RESULTS}

\subsection{Dynamical evolution of star-forming NSCs\label{sub:The-evolution-of-NSCs}}

\begin{figure*}[t]
\includegraphics[scale=0.43]{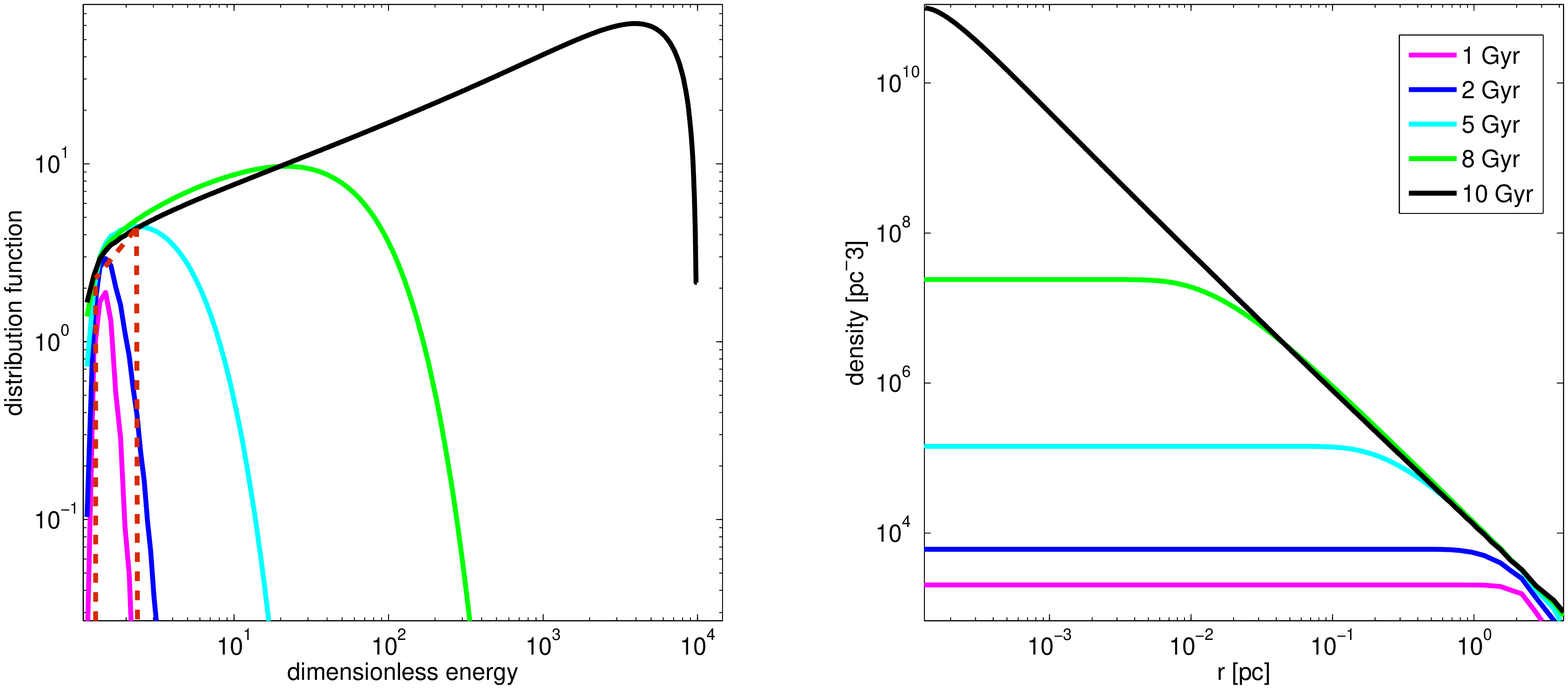}\caption{The evolution of in-situ star formation NSC at different epochs\textbf{
}taken from scenario 1. Left: the distribution function vs. the energy.
Right: the number density profile vs. the distance from the MBH. The
dashed red line presents the burst of star formation in the distribution-energy
space.\label{evolution_plot}}
\end{figure*}

As described in the previous sections, we have followed the evolution
of multiple stellar populations formed at different epochs. Fig \ref{evolution_plot}
demonstrates (model 7 in table \ref{tab:Initial-structures,-SF})
the build-up and structure evolution of an NSC which grows through
a continuous long-term in-situ star formation. The final configuration
of this NSC is very similar to that of a steady state BW cusp, and
the number densities are comparable to those observed in the GC. However,
in the BW77 scenario stars flow from the external region of the NSC
(i.e. from the background ``thermal bath'') into the the central
regions; in the SF scenario, the stars are formed in the star-forming
region, and slowly diffuse away to produce a flow which can be reversed
in direction compared with the flow in the BW77 model. The flow direction
is important when discussing populations formed at different ages;
since older stars had more time to diffuse they could be observed
farther away from the SF region compared with younger stars, and the
general direction of the stellar flow would determine the differential
distribution of older vs. younger star (e.g. an inflow would lead
to older stars dominating the inner regions, while an outflow will
lead to older stars dominating the outer region). 

\begin{figure*}[t]
\includegraphics[scale=0.55]{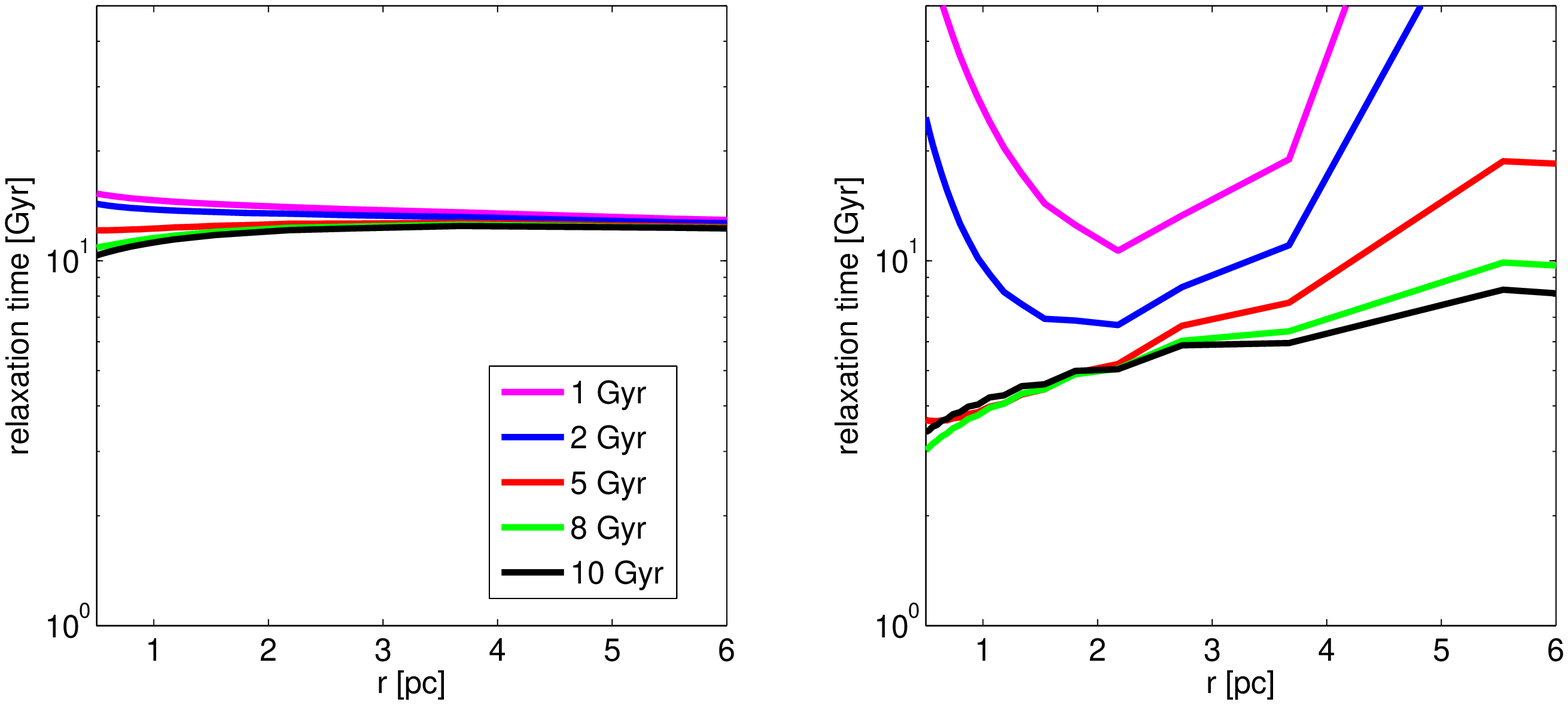}

\caption{Relaxation time vs. the distance from the MBH at different epochs
for two optional origins of an NSC: background population evolving
through simple BW evolution evaluated from scenario 1 (left), and
a background population arising from in-situ star formation evaluated
from scenario 7 (right). \label{fig:relaxation}}
\end{figure*}

The rate at which stars diffuse in phase space and attain a steady
state depends on the relaxation time of the system. Systems with shorter
relaxation times will attain a relaxed configuration, while those
with slower relaxation times might still preserve signatures of their
initial conditions (e.g. the distribution of an unrelaxed population
of stars might be closer to that of their initial post-SF distribution).
In Fig. \ref{fig:relaxation} we show the dependence of the relaxation
time on the distance from the MBH for the different evolutionary scenarios.
We see that the behavior of the relaxation time hardly changes with
time in the case where a pre-existing BW cusp is set in place prior
to the initial evolution. In this case the background population dominates
the relaxation rate. However, in models where the NSC is built completely
by SF the initial number densities are small, and they slowly increase
as more stars are formed. This is reflected by the relaxation time
evolution, which becomes systematically shorter as the NSC is built
and becomes denser. The relaxation times converge to those seen in
the case of the initial BW cusp, once the NSC grows to comparable
masses. The in-situ built NSCs therefore have longer relaxation times
during most of their evolution and would better preserve signatures
of the SF history and structure, and differential structures of stellar
populations formed at different epochs.

\subsection{The structure of star-forming NSCs}

\begin{figure*}
\includegraphics[scale=0.45]{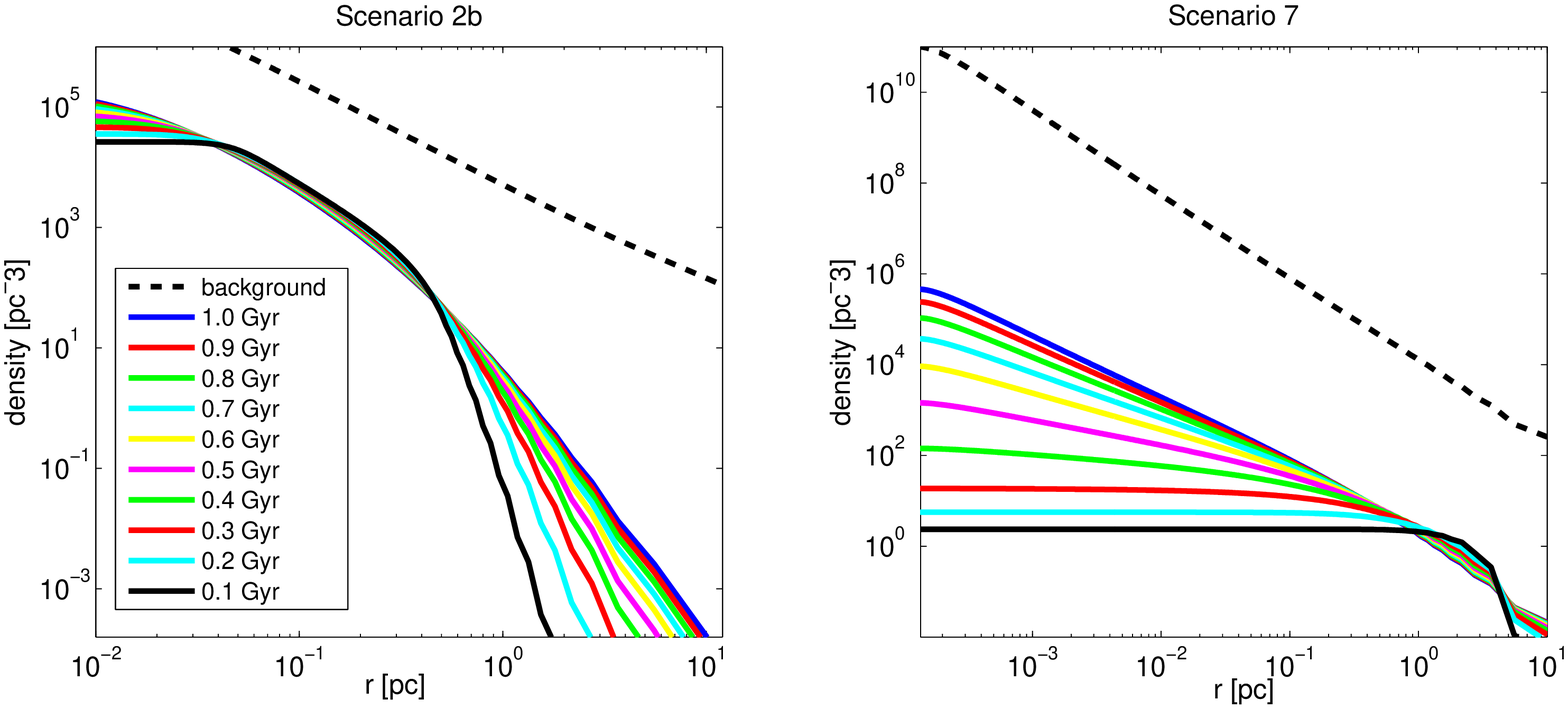}\caption{\label{fig:profile}The number density profile of a 10 Gyrs evolved
NSC with SF inside the central 2 pc (scenario 2b) and outside (Scenario
7). The number of Gyrs presents the age of the new population. The
difference in the SF range affects the final distribution of an evolved
NSC. The old background stellar population (black dashed lines) correspond
either to stars produced through in-situ SF (over the first 9 Gyr;
right) or to a pre-existing BW cusp population (left). The structures
of these old populations show an almost BW-like steady state behaviour
in both of the models, while the young stellar populations formed
in the last Gyr are not yet relaxed, and show large cores ranging
in size. These vary between $\sim0.002$ pc\textbf{ }(for the Gyr
old population) up to $\sim0.01$ pc (for the youngest 0.1 Gyr old
stellar population) in scenario 2b, and between $\sim0.01$ pc up
to $\sim0.5$ pc in scenario 7. The young populations arise from in-situ
star formation at a rate of $10^{-4}\,{\rm stars\,}yr^{-1}$, during
the last Gyr, where each SF episode continues for 100 Myrs. Detailed
model descriptions can be found in Tables \ref{tab:Initial-structures,-SF}
and \ref{tab:results}. }
\end{figure*}

\begin{figure*}[h]
\includegraphics[scale=0.75]{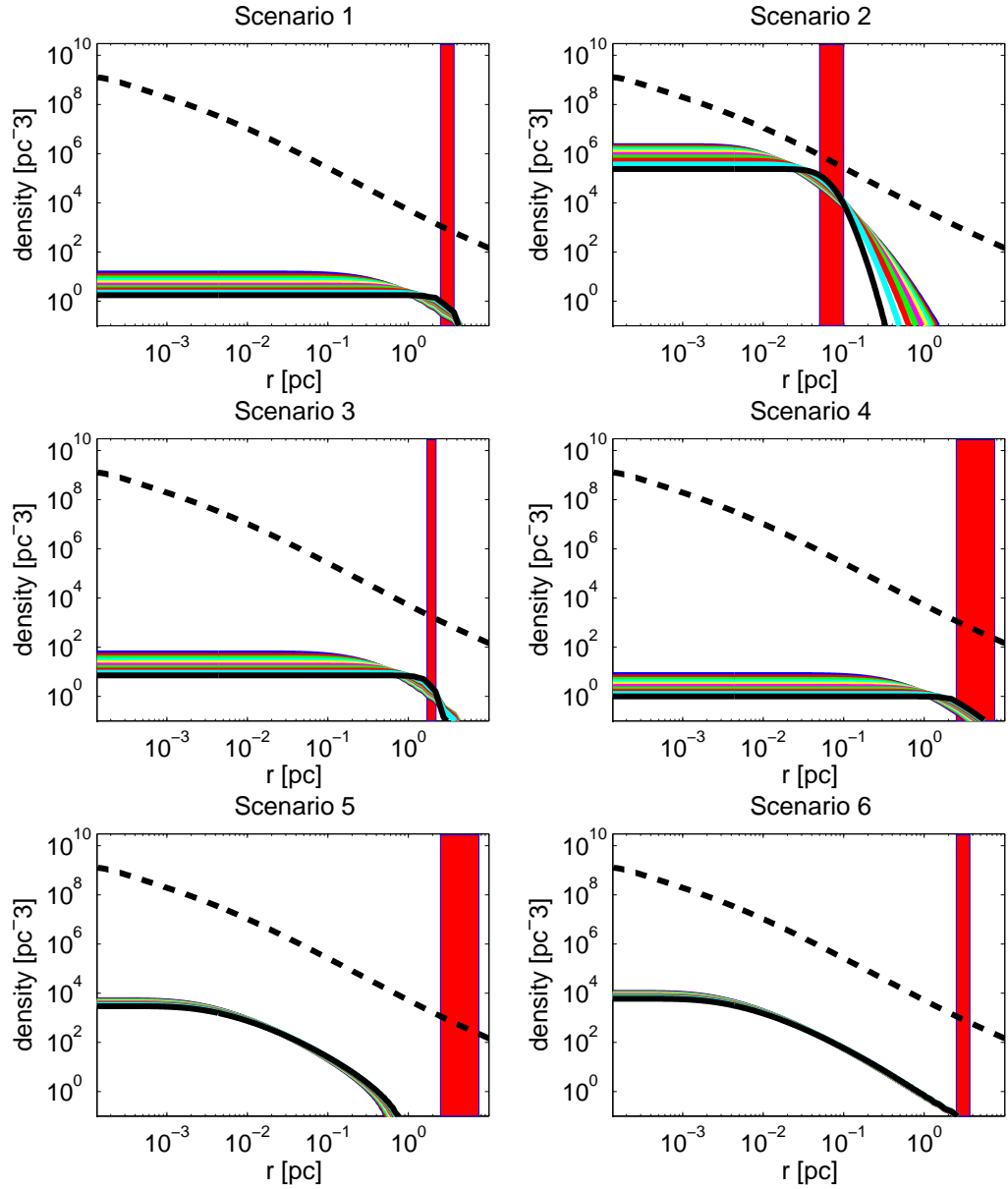}\caption{The number density profiles of NSCs with a pre-existing BW cusp and
different SF histories, after 10 Gyrs of evolution (similar to Fig.
\ref{fig:profile}; now shown for scenarios 1-6). The red shaded area
correspond to the spatial range of the SF\textbf{. }Detailed model
descriptions can be found in Tables \ref{tab:Initial-structures,-SF}
and \ref{tab:results}.\label{fig:scn1-6}}
\end{figure*}

\begin{figure*}
\includegraphics[scale=0.75]{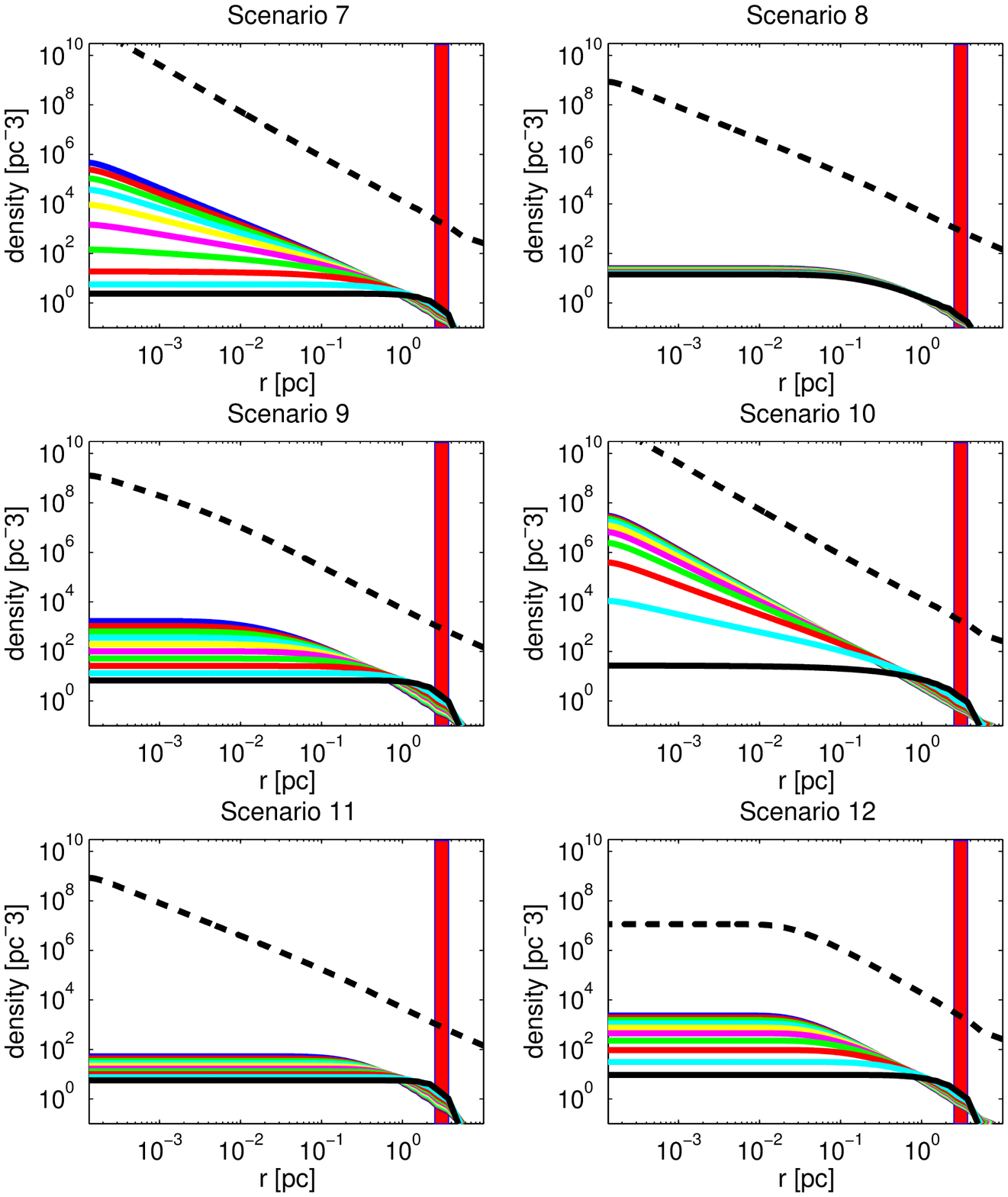}\caption{\label{fig:scn7-12} The number density profiles of NSCs with different
SF histories (similar to Fig. \ref{fig:profile}; now shown for scenarios
7-12).\textbf{ }Models 8,9 and 11 show cases of NSC evolution with
a pre-existing BW cusp (similar to models 1-6 in Fig. \ref{fig:scn1-6},
but with a different SF histories). Models 7, 10 and 12 show models
of NSCs built completely from in-situ SF (no background population).
Detailed model descriptions can be found in Tables \ref{tab:Initial-structures,-SF}
and \ref{tab:results}.}
\end{figure*}

As summarized in Tables \ref{tab:Initial-structures,-SF} and \ref{tab:results},
we have explored various models for NSCs and their SF histories. The
final structures of these NSCs after 10 Gyrs of evolution are summarized
in table \ref{tab:results} and in Figs. \ref{fig:profile}, \ref{fig:scn1-6}
and \ref{fig:scn7-12}. The four right columns of table \ref{tab:results}
summarize some of the mean properties of the clusters, including the
range of core sizes observed (i.e. where a power-law number density
profiles break to become significantly shallower) for the young stellar
populations, the total number of stars in the NSC (up to 2 pc from
the MBH) and the NSC relaxation time (at 2 pc from the MBH). The relaxation
time is calculated according to equation \ref{eq:relaxation}. 

Figs. \ref{fig:scn1-6} and \ref{fig:scn7-12} show the detailed number
density profile of the stellar populations in each cluster. Many of
these models show the existence of a core-like structure for the young
stellar populations, where the cores vary in size, and are systematically
bigger for younger populations. This can be seen in detail in Fig.
\ref{fig:profile}; similar plots are shown for the other models in
Figs. \ref{fig:scn1-6}-\ref{fig:scn7-12}. 

\section{DISCUSSION}

In this work we studied the formation and evolution of nuclear star
clusters, while taking into account the effects of in-situ star-formation
using Fokker-Plank modeling. We explored two types of scenarios: (1)
pre-existing NSCs with a BW-like structure that experience later SF
and (2) NSCs built-up completely from in-situ SF. Both type of models
assume that several epochs of gas-infall into the nuclear region triggered
SF, transforming the infalling gas into newly born stars (e.g. following
\citealp{1982A&A...105..342L}). The specific properties of such SF
epoch are unknown, and various models for SF histories, rates and
spatial characteristics have been explored. In the following we discuss
the outcomes of the evolution of such NSCs.

\subsection{The build up of nuclear stellar clusters through in-situ star formation\label{sub:The-build-up}}

As can be seen in Fig. \ref{fig:scn7-12}, NSCs built-up from in-situ
SF (models 7, 10 and 12) give rise to NSCs dominated by the stellar
population formed at earlier stages (first few Gyrs). The structure
of the older population is very similar to a steady-state BW-cusp,
and the total number of stars is comparable to that inferred for the
GC NSC. Models with significantly lower rates of SF (not shown) can
not, by themselves, reproduce number densities comparable to the NSC,
and can only form a low mass NSC with a large core (for SF outside
the central 2 pcs) or a very compact low mass NSC (for SF inside the
central 2 pc - model 2b). Conversely, models with high SF rates ($\ge10^{-3}$yr$^{-1}$),
give rise to high stellar densities, and the resulting shortening
of the relaxation times lead to the formation of an NSC and its fast
relaxation into a steady state BW-like configuration, very similar
to that obtained in the absence of SF. However, a significant negative
feedback takes place in this case, and the typical inflow of stars
in the BW-like models is replaced by an outflow, and though the number
densities of such NSCs are higher than BW-like models, they only grow
by a factor of a few, even where extreme (likely unrealistic) models
of high SF rates ($10^{-2}$$yr{}^{-1}$) are tested (not shown). 

We also note that when lower mass stellar populations (models 11-12)
are assumed (e.g. if different initial mass function are considered)
the relaxation times become longer, as expected, and late-formed younger
populations are far from achieving a steady state structure, producing
larger core-like structures, as discussed in more details below.

\subsection{Age segregation\label{sub:Age-segregation}}

\begin{figure*}[!t]
\includegraphics[scale=0.76]{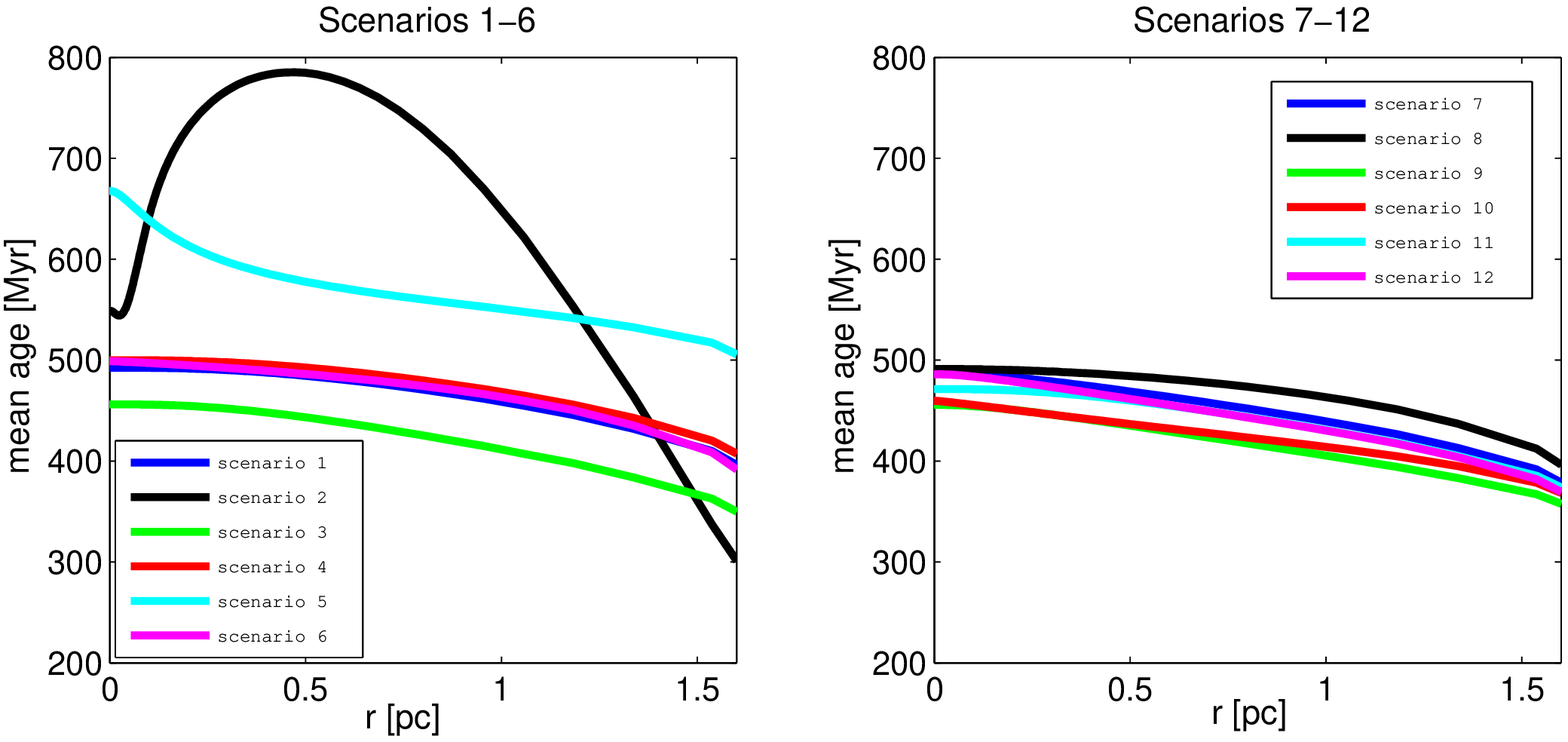}

\caption{\textbf{\label{fig:mean_age}}The mean age of the stars in the central
2 pc of an NSC, as a function of their separation from the center.
The distributions are shown at the end of the simulation for all the
scenarios (1-12). The typical 1$\sigma$ deviation from the mean is
$200$ Myrs. As can be seen, all the models predict some level of
age-segregation. While the age-segregation differences between some
of the scenarios (especially scenarios 7-12; right panel) are likely
to be too small to be observable, the different behaviour seen for
the models on the left panel is much more significant. }
\end{figure*}

The study of multiple stellar populations formed at different epochs
allow us to study the distribution of stellar ages in NSCs. In particular,
many of the models studied here show a differential structure of different
age populations. Older stellar populations are more relaxed than younger
ones, and therefore present qualitatively and quantitatively different
number density profiles. Older populations of ages comparable with
the cluster relaxation time achieve a BW-like cuspy steady state,
while younger populations behave differently. Young populations in
models with SF at the outer regions of the NSCs present large cores
in their inner number density profile, with younger populations showing
larger cores (see Figs. \ref{fig:profile}, \ref{fig:scn1-6} and
\ref{fig:scn7-12}. Given the long relaxation times in some of the
NSC models, the relaxation and slow diffusion of stars from the outer
regions to the inner ones can take longer than the age of these stellar
populations, leading to low stellar densities in the inner regions
and the core structure. Conversely, scenarios in which SF occurs in
the inner regions show, in some cases, a significant differentiation
of stellar population in the outer regions, effectively producing
an age segregated structure; different regions in the NSC outskirts
are dominated by populations of different ages, with the oldest populations
outside and younger populations closer in. These different behaviors
are well demonstrated by Fig. \ref{fig:profile}.\textbf{ }A closer
examination of the age segregation in the NSCs is shown in Fig \ref{fig:mean_age},
where the mean age of the newly formed stars (the 1-3 Gyrs old population,
depending on the model and not including the older backgro\textcolor{black}{und
population) is shown as a function of the distance from the MBH. The
evaluated standard deviation from the mean age for these scenarios
is approximately constant $\Delta_{age}\sim200$ Myr. The observed
age difference as a function of the separation from the center can
be potentially observable. As can be seen in Fig. \ref{fig:mean_age},
some of the different models show distinct age-segregation behavior
(especially scenarios 1-6), with both positive and negative age gradients.
This suggests that detailed modeling of the observed age distribution
of the stellar populations in the GC can provide a handle on the relevant
SF models. It should be noted that younger stellar populations are
more massive (higher turn-off mass), and therefore there could be
a degeneracy between interpretation of observations as arising from
mass segregation vs. age segregation. Indeed, there are clues for
the existence of extreme mass-segregation in the GC which could be
difficult to explain in the context of mass-segregation processes
\citet{2007arXiv0708.0688A}; age segregation processes may therefore
help to better understand this issue. }

It is interesting to note that the cluster-infall scenario can also
lead to age segregated NSCs, as discussed in detail by \citet{2014ApJ...784L..44P},
and the age distribution of stellar populations in NSCs can help study
not only their SF history, but potentially the impact of both the
in-situ SF and the cluster-infall scenario roles in the build-up of
NSC. However, the contribution of both processes may also give rise
to a complex age-structure of NSC. The behavior of NSCs built from
the combination of both processes is beyond the scope of this paper
and will be discussed elsewhere.

\subsection{Core-cusp structure}

As discussed above, we find that after a $\sim$10 Gyr of NSC evolution,
the younger stellar populations in NSCs may evolve to a core-like
distribution, while older population already become progressively
cuspier. These results are of great interest in light of the recent
findings about the structure of the Galactic center NSC. 

Findings by several different groups have provided strong evidence
for a core-like density profile for red-giants observed in the GC
\citet{buc+09,2009ApJ...703.1323D,2010ApJ...708..834B,2010RvMP...82.3121G}.
Though the exact size of this core is still debated, it extends throughout
the central few 0.1 pcs, with some three dimensional modeling suggesting
a core as large as 1 pc \citep{2013ApJ...779L...6D}. As noted by
\citet{2010ApJ...718..739M}, cores are ubiquitous components of galaxies
with MBHs, at least in galaxies that are bright enough or near enough
for parsec-scale features to be resolved \citep{2006ApJS..164..334F}.
In the GC, observational biases limit the study of number density
profiles only to the most luminous stars, and therefore the number
density profile of main-sequence low mass stars is still unknown.
Whether the core-like structure represents the overall distribution
of stars in the GC or only that of evolved red-giants stars is therefore
still an open question, with some important implications. Such core-like
density profiles in the central part of the GC would give rise to
long relaxation times and slow dynamical evolution. The combined effect
of slow relaxation as well as low stellar densities would be a low
rate of tidal disruption events, as well as low rate of extreme mass-ratio
gravitational-wave inspirals (EMRIs). In particular, if the GC is
representative of the structure of typical NSCs around low mass MBHs,
the total event rates from such MBH-star strong interactions could
be significantly lower than that expected from NSCs with a BW-cusp
distribution.

Several models were suggested to explain the origin of the GC core.
Some suggest it only represents the distribution of red giants, which
were preferentially destroyed through stellar collisions \citep{1996ApJ...472..153G,1999ApJ...527..835A,1999MNRAS.308..257B,2009MNRAS.393.1016D,2011ASPC..439..212D},
and/or interaction with clumps in a gaseous disk \citep{2014ApJ...781L..18A}.
However, the expected realistic collision rates are far too low to
explain a core distribution much larger than $\sim0.01$ pc \citep{2009MNRAS.393.1016D,2010ApJ...718..739M}.
The model for collisions with clumps in gaseous disk \citep{2014ApJ...781L..18A}
is extremely dependent on the distance from the MBH and assumptions
on the radial number density profile of the gas, and may therefore
potentially explain only small cores ($<$0.1 pc; compared with the
0.5-1 pc cores inferred from 3D modeling of the observations) under
specific assumptions. 

Another possibility is that the observed core distribution represents
all stars, not only the red giants. \citet{2012ApJ...750..111A} and
\citet{2014ApJ...784L..44P} have shown the the cluster-infall scenario
could lead to the formation of NSCs with significant cores. Such cores
are not limited to a specific type of stars (such as red giants),
as in the other models discussed above, but represent all the stellar
population (though age segregation could show different distributions
for stars coming from clusters infalling at different time; nevertheless,
a large difference in the core distribution of stars from different
clusters is not expected; \citealt{2014ApJ...784L..44P}).

\citet{2010ApJ...718..739M} suggested that a binary merger, or a
triaxial potential could deplete the inner regions of an NSC producing
a large core, and have shown that the long relaxation times would
not be sufficient to regrow a cusp. A similar behavior is seen in
our models, where progressively younger populations of stars formed
in the outer regions of the NSC do not relax and grow an inner cusp.
Though in both cases slow relaxation explains the non-growth of the
inner cusp, the origins of the initial core in both models differ,
and the outcomes could significantly differ as well. In particular,
the SF models studied here suggest that cores of different sizes could
exist for different stellar populations, and in particular an NSC
can have both a cusp distribution of old stars and a core distribution
for young and intermediate age stars. 

\begin{figure}[h]
\includegraphics[scale=0.4]{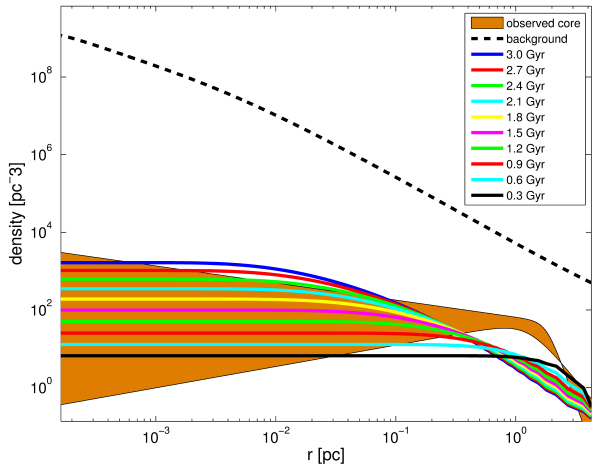}\caption{\label{fig:core}The number density profile of the GC nuclear cluster
after 10 Gyr evolution (and total star formation epoch of 3 Gyr) compared
to the inferred range of number density profiles from 3D modeling
of GC observations \citep{2013ApJ...779L...6D}. The number of Gyrs
presents the age of the new population}
\label{obs vs theory}
\end{figure}

We note that younger, more massive red giants could be more luminous
and more easily detected in observations \citep{2007ApJ...669.1024M,2011ApJ...741..108P}.
We therefore hypothesize that if such younger red-giants (up to 2-3
Gyrs old) are overly represented in observations then the observed
core could be limited to these younger populations, while the underlying
population of older stars might still have a cusp distribution. This
can be well demonstrated both in Fig. \ref{fig:scn7-12} and in more
detail in Fig. \ref{fig:core} where the model results are compared
with the 3D modeled number density profile determined by \citet{2013ApJ...779L...6D}
based on observations. As can be seen, in some models a large, pc
size core of up to a few Gyrs old stellar populations can exist. Such
a core might be consistent with the density profiles inferred from
observations, while the old stellar population preserves a typical
BW-like cusp profile. An interesting basic prediction of such models
would therefore be for stars of different ages to show a different
positions for the break in their distribution, between an inner cusp
and an outer core. A deeper study of the observational properties
and stellar evolution of populations in such models of SF origin for
the core is to be done elsewhere. 

\section{Summary}

In this work we explore the formation and evolution of NSCs arising
from in-situ star formation. We solve the Fokker-Planck equation with
additional source term accounting for the in-situ star formation,
and explore the dynamics and structure evolution of the modeled star-forming
NSCs. Our models follow the dynamics of several stellar populations
near an MBH of $4\times10^{6}M_{\odot}$ driven by 2-body relaxation
processes. We find, by comparing to observations, that the scenario
of NSCs arising from in-situ star formation can be proposed as a realistic
model of galactic nuclei formation. We show that the old stellar populations
have a cuspy distribution near an MBH (which corresponds to the classical
Bahcall \& Wolf model), while the younger populations behave differently,
depending on the regions where star-formation occurs. Stars formed
in the NSC outskirts tend to distribute in a core-like distribution,
where the size of the core decreases with time, with different stellar
population potentially showing different number density profiles.
In particular, younger stellar populations may have larger cores.
We explore such age segregation in NSCs, and present the age gradient
produced through the evolution of several population arising from
star formation at different epochs. We find that our results might
explain the origin of the core-like distribution of red giants in
the galactic center, and suggest that it might be limited to intermediate
age stellar populations ($<$3 Gyrs old), while the underlying old
stellar population might have a cuspy structure.

\acknowledgements{}

We would like to thank Clovis Hopman the use of the basic components
in his FP code for developing the FP code used in our simulations.
We acknowledge support from the I-CORE Program of the Planning and
Budgeting Committee and The Israel Science Foundation grant 1829/12.
HBP is a Deloro and BIKURA fellow.

\bibliographystyle{plainnat}

\end{document}